\documentclass[12pt]{iopart}

\def\be{\begin{equation}}
\def\te{\end{equation}}
\def\ee{\end{equation}}
\def\ba{\begin{eqnarray}}
\def\bea{\begin{eqnarray}}

\def\tea{\end{eqnarray}}
\def\ea{\end{eqnarray}}
\def\eea{\end{eqnarray}}

\def\m{\mu}
\def\n{\nu}

\newskip\humongous \humongous=0pt plus 1000pt minus 1000pt

\newif\ifdtup

\begin{document}

\title {Intrinsic and Fundamental Decoherence:\\Issues and Problems}
\author{C.~Anastopoulos$^1$ and B. L. Hu$^2$}

\address{$^1$Department of Physics, University of Patras, 26500 Patras, Greece.}

\address{$^2$Maryland Center for Fundamental Physics and Joint Quantum Institute,\\ University of
Maryland, College Park, Maryland 20742-4111 U.S.A. and\\
Perimeter Institute for Theoretical Physics\\
31 Caroline Street North, Waterloo, Ontario N2L 2Y5 Canada}
\ead{anastop@physics.upatras.gr} \ead{blhu@umd.edu}

\date{March 21, 2008}

\begin{abstract}
We investigate the meaning of gravity-induced decoherence in quantum
theory, known as `intrinsic' or `fundamental' decoherence in the
literature. We explore a range of issues relevant to this problem,
including the meaning of modified uncertainty relations, the
interpretations of the Planck scale, the distinction between quantum
and stochastic fluctuations and the role of the time variable in
quantum mechanics. We examine the specific physical assumptions that
enter into different approaches to the subject. In particular, we
critique two representative approaches that identify time
fluctuations as the origin of intrinsic or fundamental decoherence:
one that models the fluctuations by stochastic process and one that
purports to derive decoherence from the quantum fluctuations of real
clocks.
\end{abstract}

 \maketitle

\section{Introduction}

\subsection{Quantum, Intrinsic and Gravitational Decoherence}

We begin by distinguishing between the nature of quantum, intrinsic
and gravitational decoherence, as these terms are used by different
authors in special contexts for specific purposes. \textbf{Quantum
decoherence} refers to the loss of coherence in a quantum system for
various reasons, commonly due to its interaction with an environment.
It could be formulated in different ways, through decoherent or
consistent histories \cite{conhis}  or via open-system dynamics
\cite{envdec}. Much work has been done since the beginning of the
90's that we now think we have a better understanding of this issue
(see books and reviews, e.g., in \cite{decrev}).

\textbf{Intrinsic or fundamental decoherence} refers to some
intrinsic or fundamental conditions or processes which engender
decoherence in varying degrees but universal to all quantum systems.
This could come from (or could account for) the uncertainty relation
\cite{WignerUR, Karol}, some fundamental imprecision in the measuring
devices (starting with clocks and rulers) \cite{GPP} or in the
dynamics \cite{MilIntDec}, or treating time as a statistic variable
\cite{Bonifacio}. Hereby lies the possibility of alternative theories
of quantum mechanics, such as theories based on a stochastic
Schrodinger equation. This alternative form has been proposed by many
authors, notably, Ghirardi, Rimini, Weber and Pearle \cite{GRWP}
Diosi\cite{Diosi},  Gisin \cite{Gisin}, Penrose \cite{Penrose},
Persival \cite{Persival} and Hughston \cite{Hughston} et al who
suggest gravity as the origin of the modification, and Adler
\cite{Adler} who views quantum mechanics as emergent from a more
fundamental theory. (See \cite{Diosi05} for a comparison of the
approaches by Milburn, Adler, Diosi and Penrose.) Here, the search is
mainly motivated by an uneasiness in the awkward union between
general relativity and quantum mechanics, and the context can be in
the settings of quantum gravity or at today's low energy with a
classical spacetime structure and quantum matter fields. Some
protagonists believe that gravity may be needed to make quantum
physics more complete, and since universal conditions are involved in
these investigations, gravitational decoherence is often brought up
for this purpose. For a discussion on issues of decoherence in
quantum gravity, see, e.g., \cite{AHdecQG} and references therein.

\textbf{Gravitational decoherence} refers to the effect of gravity on
the decoherence of quantum systems. In principle it also pertains to
quantum gravitational effects  operative presumably at the Planck
scale, but we separate our consideration of these two regimes so the
role of gravity will not be confused.  We save the term gravitational
decoherence to refer to gravity as described by general relativity.
For this,  even weak gravity is thought to act differently in
bringing decoherence to a quantum system than other matter fields.
Some topics explored by various authors are listed in \cite{AHdecQG}.

\section{Intrinsic Decoherence, Modified Uncertainty Principle and Clock Accuracy}

From the low-energy viewpoint, the incorporation of gravity into the
quantum description involves a theory in which the fundamental
parameters are the Planck constant $\hbar$, the speed of light $c$,
and the Newton's constant $G$ whereby the relevant length-scale
would then appear to be the Planck length $l_p = (\hbar
G/c^3)^{1/2}$. The usual interpretation is that $l_p$ defines the
scale in which quantum effects become important in the description
of gravity. A complementary interpretation (relevant to the present
discussion) is that $l_p$ defines a divergence from the predictions
of ordinary low energy physics, which should arise from the
inclusion of gravity effects.

In an alternative line of thought, the Planck length is believed to
induce a modification (or limitation) to the familiar concepts of
quantum theory. In particular, it is often suggested that $l_p$
places a limit to the measurability of lengths and thus causes  a
modification of the usual Heisenberg uncertainty relations. A
recurrent theme is that these modifications may be implemented by
small changes in the standard quantum mechanical formalism (i.e.
deformation of the canonical commutation relations, changes in the
dynamics), which could lead to effects observable even at low
energies.

In the following we have selected two prominent arguments in relation
to the Planck-scale-induced effects that suggest a modification of
quantum theory. We give a brief summary of their claims followed by
some remarks on the central issues.

\subsection{Two arguments for modified uncertainty relation}

\subsubsection{Wigner's argument extended}

Wigner's original paper \cite{WignerUR} on `fundamental imprecision'
of clocks and rulers made no reference to gravity. The argument
extended to include gravity we follow here was first proposed by
Karolyhazy \cite{Karol}. This train of thought is recently used by Ng
\cite{NgUR} to suggest relations between clocks, black holes and
quantum computation (see also \cite{NgVD}).

To measure the distance $l$ between a clock of mass $m$ and a
mirror, one sends a light signal from the clock, which is reflected
at the mirror and is then received back at the clock. If the clock's
position uncertainty is $\delta l$, after time $t = 2l /c$, the
spread in the clock's uncertainty has increased by a factor of
$\hbar t/m \delta l $ due to momentum uncertainty,  so that the
total spread is $\delta l + \frac{\hbar l}{m \delta l c}$. This has
a minimum of
\begin{eqnarray}
\delta l^2 \geq \frac{\hbar l}{ m c},
\end{eqnarray}
  which decreases with mass. Hence, from this argument, heavier
clocks are more accurate.

On the other hand, the uncertainty  $\delta l$ is at least as large
as the size of the clock, which is constrained to be smaller than
the Schwarzschild radius of a body with mass $m$. Hence,
\begin{eqnarray}
\delta l \geq \frac{Gm}{c^2}.
\end{eqnarray}
Multiplying these equations one obtains
\begin{eqnarray}
\delta l \geq l_p (l/l_p)^{1/3}.
\end{eqnarray}

\subsubsection{Heisenberg microscope}

One considers the standard Heisenberg microscope argument for the
uncertainty relation, but takes into account the gravitational
interaction between a measured particle and an EM field. A photon of
energy $E$ will act on the electron with a force of the order
$\frac{GE}{c^2 L^2}$, where $L$ is the typical size of the cavity.
The characteristic time of the interaction is $L/c$, hence the force
will change the electron's  position by a factor of
\begin{eqnarray}
\delta x = \frac{1}{m} \frac{GE}{c^2 L^2} \left(\frac{L}{c}\right)^2
= \frac{GE}{m c^4}.
\end{eqnarray}
The photon's momentum is $E/c$ and this equals the uncertainty
$\Delta p$ in the determination of the electron's momentum. We
therefore have an uncertainty $\delta x = \frac{G}{mc^3} \Delta p$.
Adding this to the inherent quantum uncertainty yields
\begin{eqnarray}
\Delta x \geq \min \{\frac{\hbar}{\Delta p}, \frac{G}{mc^3} \Delta
p\} = \min \{\frac{\hbar}{\Delta p}, l_p^2 \frac{\Delta p}{\hbar}\}
\geq l_p.
\end{eqnarray}
This argument can be made more precise, using linearized general
relativity \cite{AdSan}.

\subsection{Comments}

Note that the two arguments above lead to rather different
expressions. The first argument provides a non-trivial bound in how
accurately a length can be measured -- greater lengths have higher
intrinsic uncertainty. The second provides a modification to the
uncertainty relation.

Both arguments above have features common to all approaches to this
subject matter. The estimates for uncertainty are based on separate
arguments that involve different length scales: the atomic scale for
the use of the uncertainty principle and the macroscopic scale for
the black hole argument. These results are then extrapolated to a
regime where they are both assumed to be valid jointly: the Planck
length then appears for dimensional reasons. The protagonist would
argue that it is the best one can do in order to obtain at least a
rough estimate of the consequences of a quantum theory of gravity
from the low energy viewpoint. However, such extrapolation involves
a leap of faith, our present state of knowledge being limited to low
energy phenomena.

We urge the exercise of caution in indiscriminately extending the
physics of low energies (compared to the Planck scale) to unknown
domains.  They usually involve specific {\em physical} assumptions,
which may or may not be valid. We mention two aspects:

\subsubsection{Problems in using macroscopic measurement
devices to infer microscopic physics}

Both arguments are developed with reference to the process of quantum
measurement. {\em A priori}, it is not clear  that this language can
be applied unreservedly to all length scales. Our measurement
devices, such as clocks, are typically macroscopic or mesoscopic
physical systems, which consist of a large number of microscopic
particles. Taken at face value, the arguments above seem to imply
that the notion of measurement devices makes sense at arbitrarily
small length scales: in both cases a detection of photons (either
scattered by the atoms or absorbed by the clock) is involved.
However, the measurement devices that can achieve this task usually
consist themselves of atoms and the atomic scale places strong
constraints on the accuracy of measurement.

In the first example, the accuracy in measurement of position cannot
be finer than the size of the smallest constituents of the device. In
particular, the frequency $\omega$ of the detected light signal
induces an uncertainty  $\delta t = 1/ \omega$ for the time of
detection. However, $\omega$ cannot be taken arbitrarily large in
known physics: if $\omega^{-1}$ is of the order of the plasma
frequency of the atomic electrons in the coating, then the mirror is
nonreflective and the thought experiments involving such devices lose
their meaning.

In the second example, there is a more fundamental limitation placed
by relativistic quantum physics. The minimal resolution for the
measurement of position by photons is the de Broglie wavelength of
the electron $\hbar/m_e c$, which corresponds to the energy scale
$2m_e c^2$ of pair creation for the lightest particles interacting
with the electromagnetic field. At such energies, the notion of the
Heisenberg microsope becomes problematic and a quantum field
theoretic treatment of the problem is mandatory.

\subsubsection{The Risk of violating stringent consistency conditions in Planck
scale physics}

The second example also involves a comparison of the gravity-induced
and of the familiar quantum uncertainty. While at the present state
of knowledge these sources of uncertainty are independent, this
independence is not guaranteed to hold at the Planck scale where
quantum and gravitational theories are expected to be unified or
interlinked. Even with some degree of linkage, certain
self-consistency requirement ought to exist and be respected which
may work against simple extrapolations from low energy
phenomenology. In fact, there is at least one example, in which the
justification of the uncertainty principle employs arguments from
gravity (Bohr's refutation of Einstein's box experiment)\cite{ZuWh}:
in this case the gravity uncertainties are employed in order to
justify the quantum uncertainty relation.

In essence, most proposals of this nature starting from low energy
physics and extrapolating their predictions to Planck scale regime
involve some selective processes.  They focus on some favored
(usually the easy) aspect but leave out the potential limitations
from other known physics which could exert equal or more stringent
bounds on the validity of these arguments. The reasoning which
follows from them is usually incomplete or inconsistent,  which makes
the interpretation (usually grounded on measurements at low energies)
 problematic.


\subsection{Uncertainty and noise}

We argued before that the derivation of gravity-modified uncertainty
in terms of measurement inaccuracies is not straightforward, as it
extrapolates results of quantum measurement theory at low energies
into a new regime, where habitual concepts and ordinary tools may
fail to apply. We note that the most concrete understanding of
quantum uncertainties is as mean deviations in the statistics of
measurements. Any other viewpoint  (for instance, thinking of
uncertainties as referring to the definability of the physical
magnitudes involved) is more elusive and it depends on one's
preferred interpretation of quantum mechanics.

But even if we separate the quantum uncertainties from a measurement
process,  the existence of such uncertainties does not say anything
about the physics they encode. In particular,
the statistical theory one employs for its description is a modal
simplification.  By this we mean, that we have no {\em a priori}
justification in modeling these uncertainties by stochastic processes
which are intrinsically classical. The stochastic assumption may only
be valid for uncertainties in measurements {\it classical} in nature
i.e., containing randomness at the macroscopic scale. In contrast,
quantum uncertainties are different in nature -- they cannot be
modeled by a stochastic measure as they involve non-localities and
correlations with no analogues in classical probability. See
\cite{AHdecQG} for some discussions about the difference between
quantum and stochastic effects pertaining to decoherence.

\subsection{Limitations in using particle as a field probe}

In the above discussions we urge a closer scrutiny on the physical
assumptions involved in any interpretation of new Planck scale
effects from the low energy viewpoint. Clearly, there {\em are}
physical effects associated with the introduction of the Planck
scale, but it is not straightforward to infer how they will be
manifested in the low-energy regime. In particular, one should be a
little wary whether a classical stochastic description of
uncertainties  does justice to the full quantum features of the
variables involved.

To see this we consider an analogous situation related to the
measurement of the quantum electromagnetic fields by test probes of
mass $m$ and charge $q$. The classic analysis of Bohr and Rosenfeld
\cite{BR} demonstrates that there is an uncertainty $\Delta F$ in
the determination of the electromagnetic field (EMF) strength $F$
given by
\begin{eqnarray}
\Delta F l^3 \geq \frac{q\hbar}{mc},
\end{eqnarray}
where $l$ is the typical length scale of the interaction region and
the following conditions must also hold
\begin{eqnarray}
l m c\geq \hbar \hspace{3cm} l m \geq q^2 c^2.
\end{eqnarray}

The quantity $\Delta F l q$ is a measure of the uncertainty in the
electromagnetic energy $\epsilon_{el}$ of the particle. Since the
particle description is inadequate if $\epsilon_{el} > mc^2$, we
obtain the condition
\begin{eqnarray}
l \geq \left(\frac{\hbar}{c^3}\right)^{1/2} \frac{q}{m}.
\label{limit}
\end{eqnarray}

This sets a scale to the size of the domain in which the particle
can identify a classical configuration for the EMF. For smaller
regions, the interaction between particle and field is fully quantum
and it is impossible to make a sharp distinction between them. In
gravity the role of $F$ is played by the connection $\Gamma$ and of
$\sqrt{\hbar/c^3} q$ by $l_p m$. Hence, the limit (\ref{limit})
becomes $l
> l_p$. Transferring the reasoning from the electromagnetic case,
one would say that $l_p$ determines the scale in which the coupling
of the particle to the gravitational field must take into account
the full quantum nature of the latter. In this reasoning, it does
not imply anything about the structure of spacetime or of a
necessary modification in the dynamics of low energy, but is a
statement of the fact that at $l_p$ a particle fails to detect the
classical aspects of the gravitational field's character.

The point we want to make is that the limitations posed by the
Planck length are not {\em a priori} different from those placed by
the scale $\sqrt{\hbar/c^3} q/m$ in quantum electrodynamics: At this
scale quantum field effects are strong. In electrodynamics the
fluctuations from these effects are fully quantum. Any effect they
cause at low energy is also inherently quantum: there is no
justification in treating them like classical fluctuations (i.e.,
described by a stochastic process), unless we assume a specific
regime for the quantum field. In particular, the effects of the
fluctuations of the EMF at low energies ($E<<mc^2$) have been
studied in \cite{BaCa, HM2, AnZ, JH, LH07}. There, it was shown that
the `noise' induced by these fluctuations is non-Markovian and does
not cause significant decoherence effects in the microscopic regime.
In other words, the coherence of the EMF vacuum does not allow for
the {\em a priori} generation of classical (i.e., decohering)
fluctuations in the quantum motion of the particle.


\section{ Intrinsic Decoherence from Time Fluctuations}

\subsection{Time fluctuations}
A key theme in the approaches to intrinsic decoherence is that it is
induced by the fluctuations or uncertainties in time caused by
gravity. This idea originates with Penrose \cite{Penrose}, who noted
that a superposition of two states with different mass density
distributions would lead to a superposition of two different
spacetimes, and hence an ambiguity in the notion of the time
variable, through which the superimposed quantum states evolved. He
then suggested that this ambiguity might imply that such
superpositions are not stable and thus decay. This would suggest
that (in the non-relativistic regime) in effect there is decoherence
in the `basis' of energy density. A special case, if the
superimposed states have support in the same region of space, the
decoherence would be in the energy basis.

The argument above is stated using only the basic principles of
quantum theory and general relativity. Without further assumptions
it only suggests the plausibility of an intrinsic decoherence effect
and an order of magnitude estimate in the non-relativistic regime.
However, if one makes additional assumptions about the nature of the
time fluctuations, it is possible to derive a master equation for
the quantum state and thereby study decoherence as a dynamical
effect.

The additional assumptions usually refer to the statistical
properties of the time fluctuations. For example, Diosi has
considered that the fluctuations of time are expressed in terms of
fluctuations of the Newtonian potential, which can be seen as a
random field satisfying a Markovian process \cite{Diosi}.

\subsection{Milburn's Intrinsic Decoherence}

We  next examine one representative scheme of intrinsic decoherence,
that proposed by Milburn \cite{MilIntDec}. Actually, Milburn's
discussion is in a context more general than that of gravitational
induced fluctuations. He studies the effects of randomness in the
preparation of experiments performed on relativistic systems. He
assumes that there is a (perhaps fundamental) minimal fluctuation
$\epsilon$ in the determination of some physical quantity $A$. Let
$\hat{g}$ be the generator corresponding to translations in the
value of $A$ (e.g., the Hamiltonian if $A$ is time, the momentum if
$A$ is a spatial displacement). Assuming that the fluctuations are
described by classical probability theory, there is a probability
$p_n(\epsilon, \theta)$ that  $n$ random shifts of order $\epsilon$
will result in a change of the parameter $A$ from $A = 0$ to $A =
\theta$. Then one can write the following equation for the density
matrix
\begin{eqnarray}
\rho_{\epsilon}(\theta) = \sum_{n=0}^{\infty} p_n(\epsilon, \theta)
e^{-i n \epsilon \hat{g}}\rho(0) e^{i n \epsilon \hat{g}},
\end{eqnarray}
with the condition that for $\epsilon = 0$ this law corresponds to a
unitary transformation
\begin{eqnarray}
\lim_{\epsilon \rightarrow 0} \rho_{\epsilon}(\theta) = e^{-i \theta
\hat{g}}\rho(0) e^{i \theta \hat{g}}.
\end{eqnarray}

The equation above defines a positive semigroup
\cite{Lindblad,Alicki}. There is substantial freedom in its choice,
as it depends on the determination of the probability function
$p_n$. A convenient choice  is the Poisson distribution
\begin{eqnarray}
p_n(\epsilon, \theta) = \frac{(\theta/\epsilon)^n}{n!} e^{-
\theta/\epsilon},
\end{eqnarray}
which leads to a semigroup equation
\begin{eqnarray}
\frac{d \rho}{d \theta} = \frac{1}{\epsilon} \left( e^{-i \epsilon
\hat{g}} \rho e^{i \epsilon \hat{g}} - \rho \right).
\end{eqnarray}

Milburn then explores this semi-group transformation for the special
case of time-translations ($\hat{g} = \hat{H}$) and at displacements
($\hat{g} = \hat{P}_i$), in some spatial direction labeled by $i$.
This choice allows for the preservation of energy and momentum, He
is mainly interested in uncertainty relations. In the first case
(and assuming a specific scheme for time measurements) he finds an
exponential degradation of the accuracy: $\delta t \sim e^{ a t}$,
in terms of some state-dependent factor $a$.

In the second case, he finds (for an initial Gaussian state)
\begin{eqnarray}
(\delta X)^2 (\Delta p)_0^2 \geq \frac{1}{4} + \epsilon X (\Delta
p)_0^2,
\end{eqnarray}
where $X$ is the particle's distance from the origin, $\delta X$ the
position uncertainty and $(\Delta p)_0$ is the momentum uncertainty
of the initial state. He then compares this result to the analogous
expressions for the quantum-gravity-modified uncertainty relations,
by identifying $\epsilon$ with the Planck length. This construction
is fairly
general and it can be generalized to the relativistic regime. \\

\subsection{Comments}

Strictly speaking, Milburn's formalism deals with the effect of
incremental random effects on a quantum system. A priori, there is no
compelling reason why the Planck scale, or gravity needs to appear.
To make the connection, one identifies the minimal shifts $\epsilon$
with the fundamental uncertainties in measurement of position or time
arising from the Planck regime. However, in doing so one is making a
very strong assumption, namely, that these fluctuations are
stochastic and they can thus be modeled by a probability distribution
$p_n(\epsilon, \theta)$.

However, quantum effects cannot in general be described in terms of
stochastic processes. Coherence and nonlocality (e.g., Bell's
theorem) are inherent quantum properties which are lost in a
stochastic description. Quantum theory involves `interference'
effects which do not allow the definition of a stochastic measure.
Milburn' s assumption inconspicuously treats the Planck-scale
fluctuations, what he calls the `quantum' foam,  like fluctuations
observed in the macroscopic world. It also ignores possible quantum
correlations between the quantum system and its environment which is
fundamentally also quantum in nature. Only after coarse-graining
(often rather drastically) could it begin to show some classical
stochastic features. Interchanging quantum fluctuations with
classical noise loses precious quantum information which changes the
character of the quantum system fundamentally.

This is particularly pertinent with fluctuations that refer to time.
Time occupies a special position in quantum theory: there is no
natural time operator in quantum mechanics, and the construction of
probabilities for time is rather intricate as even elementary
problems like the time-of-arrival indicate. The notion of {\em
quantum} temporal fluctuations is therefore both technically and
conceptually different from fluctuations of usual observables (see
\cite{timemuga} and references therein, also \cite{AnSav06}).

Concerning possible modifications in the uncertainty relation, we
note that such modifications are typically generated by non-unitary
dynamics. For example, they have been studied in models like quantum
Brownian motion \cite{AnaHalUnc,HuZhaUnc}. The growth of
uncertainties  due to noisy-dynamics is a well-established effect.
It follows from the fact that the density matrix of a pure state
becomes mixed.  The uncertainties obtained from such treatments are
not fundamentally quantum, in the sense that their origin does not
lie in the limits posed by the non-commutativity of operators, but
in the {\em degradation of predictability} induced by the external
noise.

We note that the critique above refers to any approach towards
gravitationally induced decoherence that makes specific assumptions
about the statistical behavior of the temporal fluctuations.
However,  in Penrose's proposal \cite{Penrose} such assumptions are
not made (at the price of not writing a master equation) but the
prediction of gravitationally induced decoherence proceeds mainly
from order-of-magnitude estimations and the structural
incompatibility between quantum mechanics and general relativity in
the treatment of time.

\section{Fundamental Decoherence, Quantum Mechanics with Real Clocks}

Another proposal worthy of some closer scrutiny is the so called
fundamental decoherence from quantum gravity of Gambini, Porto and
Pullin (GPP) \cite{GPP}. A distinguishing feature in this approach
is that it purports to describe a general scheme for the fully {\em
quantum} treatment of temporal fluctuations. Their argument goes
through the following steps:

\subsection{GPP's Theses}

(i) Unlike in standard quantum mechanics (QM) where the spacetime
structure is fixed, in quantum gravity one cannot rely on a time $t$
external to the system. For time evolution to make sense, it is
necessary that one introduces variables in the system's phase space
that take the role of clocks and rulers. One then has to express the
evolution law  in terms of readings for these clocks and rulers (relational time).
\\
(ii) Clocks are also quantum systems, hence they are also subject to
quantum fluctuations. Expressing time evolution in terms of measurable time
 leads to non-unitary dynamics, and therefore decoherence.
 \\
(iii) One can translate the usual Schr\"odinger evolution of standard
QM into a form, in which the temporal parameter $T$ is the reading of
the clock. Since $T$ is a quantum variable, it includes some
randomness and is not related to $t$ through a delta function.
Following a specific procedure, one can then obtain a master equation
for the density matrix, which is of the Lindblad type. Hence the use
of a non-ideal clock may generate decoherence, which GPP call
`fundamental decoherence'. The relevant
diffusion parameter is determined by the intrinsic uncertainty of the clock variable.
\\
(iv) At the Planck scale, Wigner's argument extended by Karolyhazy
and Ng (i.e. black hole creation if we try to measure space or time
with arbitrary precision) can be invoked to argue that there is a
lower limit to the possible resolution of clocks. Hence, the
arguments above apply and there has to be fundamental decoherence.

\subsection{Comments}

Assumption (i) involves specific interpretations of time in quantum
gravity -- see \cite{relational_time} and \cite{critrel} for
critique. This is not the only possible interpretation; it is however
reasonable and widely held. However, even accepting (i), step (ii)
does not necessarily follow. The reason is that there is no {\em a
priori} guarantee that the effective dynamics obtained from writing
the evolution in terms of clock time will lead to decoherent
dynamics. A quantum clock may very well exhibit strong quantum
coherence properties, which will be inherited to the evolution of the
system. In other words, it is necessary that the fluctuations of the
clock are sufficiently classical so that the effective dynamics will
be decoherent.

Hence, to support (ii) one has to demonstrate explicitly a case that
dynamics written in terms of clock time leads to decoherence.  GPP
have provided an argument for this purpose \cite{GPP}. Note that
while the argument was initially phrased in the context of quantum
gravity with discrete structure, it also holds for continuous time
and has an analogue within the context of ordinary non-relativistic
quantum mechanics. The key point in GPP's argument is a specific use
of quantum probability, as explained below.

\subsection{Key Point}

In standard QM, the time $t$ is an external parameter of the system
(in discrete gravity this role is played by the time-step $n$). Let
$T$ be the clock variable. One can then introduce projectors along
the range of $T$, say $P_{\Delta_T}$, which correspond to $T$ taking
values in an interval $\Delta_T$.

Similarly one introduces projectors $P_{\Delta_O}(t)$ corresponding
to an observable $O$ taking values in a set $\Delta_O$. The key step
in GPP's assumption is the following `expression' for the conditional
probability that ``the observable $O$ lies in $\Delta_O$ provided $T$
lies in $\Delta T$."
\begin{eqnarray}
{\cal P}[ O \in \Delta_O|T \in \Delta_T] = \lim_{\tau \rightarrow
\infty} \frac{ \int_{0}^{\tau} dt Tr[P_{\Delta_O}(t) P_{\Delta_T}(t)
\rho_0 P_{\Delta_T}(t)]}{\int_{0}^{\tau} dt Tr [\rho_0
P_{\Delta_T}(t)]}. \label{cond_prop}
\end{eqnarray}
where $P(t) = U^{\dagger}(t)PU(t)$ is the Heisenberg time evolution
of the projector $P$; $U(t) = e^{-iHt}$.

From this expression, using arguments of standard probability theory
it is easy to change variables in the density matrix $\rho(t)$
(evolved unitarily) and write it in terms of $T$. Indeed, the spread
of $T$ with respect to $t$ will be sufficient to show decoherence.
In our opinion, the main issue  is how Eq. (\ref{cond_prop}) can be
justified in the context of standard quantum theory.

\subsection{Critique}

We find two problems with this line of reasoning. First, the
interpretation of the expression (\ref{cond_prop}) as the conditional
probability that ``the observable $O$ lies in $\Delta_O$ provided $T$
lies in $\Delta T$" is not justifiable by the rules of quantum theory
and the standard calculus of probabilities. Second, even with a
justifiable expression for such probabilities, propositions such as
the above are in fact propositions about histories and are subject to
the problems of defining probabilities for quantum histories
\cite{conhis}.

\subsubsection{On the conditional probability (\ref{cond_prop})}

 If $A$ and $B$ are two events, the
conditional probability of ``$A$ provided $B$" is given by the ratio
of the joint probability for A and B to the probability for B,
provided the latter is non-zero. Then Eq. (\ref{cond_prop})  defines
a conditional probability as above if the numerator in
(\ref{cond_prop}) is proportional to the joint probability of $O$
lying in $\Delta_O$ and $T$ lying in $\Delta_T$, and the denominator
proportional to the probability that $T$ lies in $\Delta_T$ at some
time $t$. The proportionality factor must have dimension of time
inverse; the usual choice is $C(\tau) = \frac{1}{ \tau}$.

Hence, the interpretation of (\ref{cond_prop}) as a conditional
probability requires that the joint probability of $O$ lying in
$\Delta_O$ and $T$ lying in $\Delta_T$ be given by
\begin{eqnarray}
\lim_{\tau \rightarrow \infty} C(\tau) \int_{0}^{\tau} dt
Tr[P_{\Delta_O}(t) P_{\Delta_T}(t) \rho_0 P_{\Delta_T}(t)].
\end{eqnarray}
An analogous expression should hold when the time is taken discrete,
namely,
\begin{eqnarray}
\lim_{N \rightarrow \infty} C(N) \sum_{i = 1}^{N}
Tr[P_{\Delta_O}(t_i) P_{\Delta_T}(t_i) \rho_0 P_{\Delta_T}(t_i)].
\label{probdis}
\end{eqnarray}

 The question is how  the probabilistic interpretation of the
expressions above are justified in the context of standard quantum
probability. In Eq. (\ref{probdis}), the  term $p_i :=
Tr[P_{\Delta_O}(t_i) P_{\Delta_T}(t_i) \rho_0 P_{\Delta_T}(t_i)]$
can be interpreted as the probability that $O \in \Delta_O$ at time
$t_i$ and $T \in \Delta_T$ at time $t_i$\footnote{If $O$ and $T$ do
not commute, then there is the tacit assumption that the
determination of $T$ takes place just prior to the determination of
$O$, i.e. $T$ is determined at $t_i - \epsilon$ and $O$ at time
$t_i+\epsilon$, where $\epsilon$ is very small and is eventually
taken  to zero.}. Now, GPP consider the probability (\ref{probdis})
as corresponding to the statement that ``$O \in \Delta_O, T\in
\Delta_T$ at some time $t_i$". By the expression ``at some time" one
means ``either at time $t_1$, or at $t_2$, or at $t_3$ etc for all
times $t_i$." Hence, they seem to argue, it should equal the sum of
the probabilities that ``$O \in \Delta_O, T\in \Delta_T$ at time
$t_i$".

In effect, this definition relies on the following assumption
\\ \\
 ``probability of $O \in \Delta_O,
T\in \Delta_T$ at time $t$" + ``probability of $O \in \Delta_O, T\in
\Delta_T$ at time $t'$" = ``probability of $O \in \Delta_O, T\in
\Delta_T$ either at time $t$ or at time $t'$".
\\ \\
However, this assumption is erroneous. The reason is that the
additivity of probabilities holds only for {\em exclusive}
alternatives, i.e. if their join is the empty set\footnote{We remind
the reader that for two sets $A$ and $B$, $p(A \cup B) = p(A) + p(B)
- p(A \cap B)$.}. This is not the case here. To see this, one should
note that in the space of alternatives for the system, $t_i$ is {\em
not} a random variable, but a parameter of the system's histories.
This means that, unlike what is the case for random variables,
alternatives labeled by different values of $t_i$ are {\em not
necessarily} exclusive\footnote{Note that the above holds for any
parameter, discrete or continuous, that can be used as a label for
the causal ordering of the properties of the system. It does have to
be named `time' or to coincide with the time of the non-relativistic
theory.}. This follows from the `logical' structure of alternatives
involving time in quantum theory--see \cite{IL, ILSS} ofr details.
The alternatives considered here fall into the class of the
so-called spacetime coarse-grainings \cite{timeaver, AnSav06}.

An example from standard probability theory serves to illustrate the
point above. Let us consider the space of alternatives for the
position $x$ of a particle at different times. The proposition ``the
particle lies at $x = x_1$ at time $t_1$" is not disjoint to the
proposition that ``the particle lies at $x = x_1$ at time $t_2$":
clearly, there are paths that satisfy {\em both} propositions, for
example a path that satisfies $x = x_1$ at all times $t_i$. One
cannot therefore obtain the probability of their join by summing the
individual probabilities.

Eq. (\ref{probdis}) can be viewed as a time average of single-time
probabilities for $O \in \Delta_O, T\in \Delta_T$, but as we showed
this time-average is not in any sense related to the proposition
that "$O \in \Delta_O, T\in \Delta_T$ at any time $t_i$". In fact it
is not clear  that it refers to any physical statement at all.

To construct the correct expression for  the probability that ``$O
\in \Delta_O, T\in \Delta_T$ at some time $t_i$" one notes that this
proposition is the negation of the proposition that "$O \notin
\Delta_O, T \notin \Delta_T$ at {\em all} times $t_i$". Hence, the
probabilities for these two alternatives add up to unity. Now, the
probability for ``$O \notin \Delta_O, T \notin \Delta_T$ at {\em all}
times $t_i$" is given by the standard expression
\begin{eqnarray}
Tr \left[ (1 - P_{\Delta_O})(t_N) (1 - P_{\Delta_T})(t_N) \ldots (1
- P_{\Delta_O})(t_1) (1 - P_{\Delta_T})(t_1) \rho_0 \right.\nonumber \\
\left. \times (1 - P_{\Delta_O})(t_1) (1 - P_{\Delta_T})(t_1) \ldots
(1 - P_{\Delta_O})(t_N) (1 - P_{\Delta_T})(t_N) \right],
 \end{eqnarray}
where the limit $N \rightarrow \infty$ may be taken at the end. Then
the probability for the proposition ``$O \in \Delta_O, T\in \Delta_T$
at some time $t_i$" equals
\begin{eqnarray}
p(O \in \Delta_O, T\in \Delta_T) =   \nonumber \\ 1 - \lim_{N
\rightarrow \infty} Tr  \left[ \left(\prod_{i=1}^{N}(1 -
P_{\Delta_O})(t_i) (1 - P_{\Delta_T})(t_i)\right)^{\dagger}
 \rho_0 \prod_{i=1}^{N}(1 - P_{\Delta_O})(t_i) (1 -
P_{\Delta_T})(t_i) \right], \;\;\;\;\; \; \label{correct}
\end{eqnarray}
which clearly is very different from (\ref{probdis}).

\subsubsection{The issue of defining probabilities for histories}

Eq. (\ref{correct}) is obtained by standard probabilistic arguments
together with the quantum reduction rule. However, it has problems
typical to all probability assignment for histories, i.e.
propositions about a quantum system that refer to more than one
instant of time \cite{conhis}. The presumed probabilities
(\ref{correct}) -- but also the probabilities (\ref{probdis}) -- do
not satisfy the Kolmogorov additivity conditions
\begin{eqnarray}
p(O\in \Delta_O, T \in \Delta^1_{T} \cup \Delta^2_{T}) = p(O\in
\Delta_O, T \in \Delta^1_{T}) + p(O\in \Delta_O, T \in
\Delta^2_{T}), \;\;\;\;\;\;
\end{eqnarray}
for any disjoint sets $\Delta^1_T, \Delta^2_T$. This implies that
Eq. (\ref{correct}) fails to define a genuine probability measure.

The demonstration is standard \cite{conhis} and can be seen by
direct inspection of Eq. (\ref{correct}).\footnote{For Eq.
(\ref{probdis}), the probability corresponding to $O \in \Delta_O, T
\in \Delta_T^1 \cup \Delta_T^2$ can easily be shown to  differ from
the sum of the probability for $O \in \Delta_O, T \in \Delta_T^1$ to
the probability for $O \in \Delta_O, T \in  \Delta_T^2$ by an
`interference' term $2 Re \left[P_{\Delta_O}(t_i)
P_{\Delta_T^1}(t_i)  \rho_0 P_{\Delta_T^2}(t_i)  \right]$, which is
generically non-zero. Hence, Eq. (\ref{cond_prop}) also does not
define a genuine probability measure for the space of values for $O$
and $T$. }

To define a genuine probability measure for the joint values of $O$
and $T$ out of Eq. (\ref{correct}), or (\ref{probdis}), one must
either restrict to specific sets of histories that satisfy a
decoherence condition as in the consistent/decoherent histories
approach, or to modify (\ref{correct}) in a way that allows for the
definition of a probability measure in the context of measurements
-- see \cite{Ana06} for details. The former case effectively assumes
the decoherence one is supposed to find, while the latter refers to
an effectively open system \footnote{ Since the expression
effectively involves an integral over all times, one should assume a
{\em continuous-time measurement} scheme, i.e. the presence of a
device measuring (approximately) both $T$ and $O$ at all times $t$.
In other words, one would have to assume a continuous `reduction of
the wave packet', in order to suppress the generic interference
terms in the probability with respect to time.}. Either way, even
with a meaningful construction of conditional probabilities, the
domain of validity for the description of evolution in terms of
relational time is reduced substantially.

This problem is a special case of a more general fact. If we want to
treat time fluctuations within standard quantum theory, we must
effectively work within a histories scheme, because a theory with
time fluctuations necessarily involves properties of the system at
more than one `moments of time', i.e., histories. One then needs to
address the problem of adequately defining probabilities for such
histories. This is a highly non-trivial task that adds a substantial
degree of difficulty in the problem.

\section{Summary}

There is substantial motivation for the study of intrinsic or
fundamental decoherence in quantum mechanics, especially in relation
to gravitational effects. \\ \\
a. A modification of quantum mechanics at this level could provide a
resolution of long-standing problems, such as the quantum measurement problem. \\
b. The incompatibility between quantum theory and general relativity
(especially in the treatment of time) suggests a modification of
known physics in the regime in which both theories apply and  this
could have observable
consequences even in non-relativistic physics. \\
c. It is conceivable that such modifications uncover remnants of
processes fully present at much higher energy scales, such as the
Planck length.
\\ \\
The aim of such studies is to identify new physics pertaining to the
interplay between gravity and quantum theory in a different (and
observationally more accessible) regime than the one usually
attributed to quantum gravity. We point out any such theoretical
treatment may risk making strong {\em ad hoc} physical assumptions
about the nature and origin of such effects, such as intrinsic or
fundamental decoherence. One ought to be wary of the danger that
predictions drawn from such deductions only reflect the assumptions
entering in their derivations.

Here we focussed on the analysis of the basic assumptions implicit in
such schemes. We first discussed the notion of a gravity modified
uncertainty relations: the Planck scale enters explicitly into such
relations. However, this is primarily due to dimensional reasons:
derivations of such modifications involve a huge extrapolation of
known physics to an unknown regime. Moreover, the existence of
fluctuations (say, for time) says nothing about the physics they
encode. Emphatically, whether such fluctuations lead to fundamental
decoherence depends on the probabilistic theory that describes them:
we point out that fluctuations arising from a state that preserves
quantum coherence (like for example the vacuum of the electromagnetic
field) are not likely to give significant decoherence effects. The
invocation of a stochastic process modelling such fluctuations
involves {\em a strong physical requirement} that the relevant
quantum degrees of freedom have been already classicalized, or that
they are classical to begin with.

When the fluctuations refer to time the situation becomes more
complex. Setting aside any conceptual issues about the meaning of
temporal fluctuations,  their treatment is very intricate at the
quantum level, because of the distinguishing role time plays in
quantum mechanics and the fact that the corresponding probabilities
are in general not well-defined.
\\ \\
 {\bf Acknowledgment}  We thank
Albert Roura for discussions on a range of problems in gravitational
decoherence, and the referees for suggesting some new references and
correcting some of our citations. This work is supported in part by
NSF grants PHY-0601550 and PHY-0426696.
\\ \\


\begin{thebibliography}{}


\bibitem{conhis}
R. B. Griffiths, J. Stat. Phys. {\bf 36}, 219 (1984); R. Omn\'es, J.
Stat Phys. {\bf 53}, 893 (1988); {\it ibid.} {\bf 53} 933 (1988);
{\it ibid.} {\bf 53} 957 (1988); {\it ibid.} {\bf 54}, 357 (1988);
Ann. Phys. (NY) {\bf 201}, 354 (1990); Rev. Mod. Phys. {\bf 64}, 339
(1992); {\it The Interpretation of Quantum Mechanics} (Princeton UP,
Princeton, 1994); M. Gell-Mann and J. B. Hartle, in {\it Complexity,
Entropy and the Physics of Information}, ed. by W. H. Zurek
(Addison-Wesley, Reading, 1990); Phys. Rev. D {\bf 47}, 3345 (1993);
J. B. Hartle, ``Quantum Mechanics of Closed Systems'' in {\it
Directions in General Relativity} Vol. 1, eds B. L. Hu, M. P. Ryan
and C. V. Vishveswara (Cambridge Univ., Cambridge, 1993); H. F.
Dowker and J. J. Halliwell, Phys. Rev. D {\bf 46}, 1580 (1992); J.
J. Halliwell, Phys. Rev. D {\bf 48}, 4785 (1993).

\bibitem{envdec}
W. H. Zurek, Phys. Rev. D {\bf 24}, 1516 (1981); D26, 1862 (1982);
in {\it Frontiers of Nonequilibrium Statistical Physics}, ed. G. T.
Moore and M. O. Scully (Plenum, N. Y., 1986); Physics Today {\bf
44}, 36 (1991); E. Joos and H. D. Zeh, Z. Phys. B {\bf 59}, 223
(1985); A. O. Caldeira and A. J. Leggett, Phys. Rev. A {\bf 31},
1059 (1985); W. G. Unruh and W. H. Zurek, Phys. Rev. D {\bf 40},
1071 (1989); B. L. Hu, J. P. Paz and Y. Zhang, Phys. Rev. D {\bf
45}, 2843 (1992); W. H. Zurek, Prog. Theor. Phys. {\bf 89}, 281
(1993).


\bibitem{decrev}  D. Giulini {\it et
al}, {\it Decoherence and the Appearance of a Classical World in
Quantum Theory} (Springer Verlag, Berlin, 1996). J. P. Paz and W. H.
Zurek, in {\sl Coherent Atomic Matter Waves}, Les Houches Lectures,
edited by R. Kaiser, C. Westbrook, and F. David (Springer, Berlin
2001), p. 533;  W. H. Zurek,  Rev. Mod. Phys. 75, 715-775 (2003); M.
Schlosshauer, Rev. Mod. Phys. 76, 1267 (2005).


\bibitem{WignerUR} E. P. Wigner, Rev. Mod. Phys. 29, 255 (1957).

\bibitem{Karol}F. Karolyhazy, Nuovo Cim. 52, 390 (1966)



\bibitem{GPP} R. Gambini, R. Porto and J. Pullin, Phys. Rev. Lett. 93,
240401 (2004); R. Gambini, R. Porto and J. Pullin, New J. Phys. 6,
45 (2004); R. Gambini, R. Porto and J. Pullin, ``Fundamental
decoherence from quantum gravity: a pedagogical review"
[gr-qc/0603090]; R. Gambini and J. Pullin, Found. Phys. 37, 1074
(2007).

\bibitem{MilIntDec} G. J. Milburn, New J. Phys. 8, 96 (2006); Phys. Rev. A44, 5401
(1991).

\bibitem{Bonifacio} R. Bonifacio, Nuovo Cim. B114, 473-488 (1999)
[arXiv:quant-ph/9901063]

\bibitem{GRWP} G. C. Ghirardi, A. Rimini and T. Weber, Phys. Rev.
D34, 470 (1986). P. Pearle, Phys. Rev. D13, 857 (1976); 29, 235
(1984); 33, 2240 (1986). G. C. Ghirardi, P. Pearle and A. Rimini,
Phys. Rev. A42, 78 (1990); G. C. Ghirardi and A. Bassi, Phys. Rep.
379, 257 (2003).

\bibitem{Diosi} L. Diosi, Phys. Lett. A105, 199 (1984);
Phys. Lett. 120, 377 (1987); L. Diosi, Phys. Rev. A40, 1165 (1989).

\bibitem{Gisin} N. Gisin, Phys. Rev. Lett. {\bf 52}, 1657 (1984).

\bibitem{Penrose} R. Penrose, in R. Penrose and C. J. Isham editors, {\em Quantum Concepts in Space and Time}
(Oxford, 1986, Clarendon PRess);  Gen. Rel. Grav. 28, 581 (1996);
    Phil. Trans. Roy. Soc.  (London) A, 356, 1927 (1998).

\bibitem{Persival} I. C. Percival, {\sl Quantum State Diffusion}
(Cambridge University Press, 1998); Phys. World 10, 43 (1997). I. C.
Percival and W. T. Strunz, Proc. R. Soc. A 453, 431 (1997).

\bibitem{Hughston} L. Hughston, Proc. Roy. Soc. Lond. A 452, 953-979
(1996). S. L. Adler and L. P. Horwitz, J. Math. Phys. {\bf 41} 2485
(2000).

\bibitem{Adler}  S. L. Adler, {\sl Quantum Theory as an Emergent
Phenomenon} (Cambridge University Press, 2004).

\bibitem{Diosi05} L. Diosi, Braz. J. Phys. 35, 260 (2005)
[quant-ph/0412154].


\bibitem{AHdecQG} C. Anastopoulos and B. L. Hu, J. Phys. Conf. Ser. 67, 012012 (2007) [gr-qc/0703137]

\bibitem{Lindblad}  G. Lindblad, Commun. Math. Phys. 48, 119 (1976).
V. Gorini, A. Kossakowski, and E. C. G. Sudarshan, J. Math. Phys.
(N.Y.) 17, 821 (1976).

\bibitem{Alicki} R. Alicki and K. Lendi, Quantum Dynamical
Semigroups and Applications (Springer, Berlin,1987).



\bibitem{NgUR} Y. J. Ng,
Phy. Rev. Lett. 86, 2946 (2001); Erratum-ibid. 88 (2002) 139902.
``Quantum Foam and Quantum Gravity Phenomenology" [gr-qc/0405078]

\bibitem{NgVD} Y. J. Ng and H. van Dam, Ann. Phys. N.Y. Acad. Sci.
755, 579 (1995); Mod. Phys. Lett. A9, 335 (1994)



\bibitem{AdSan} R. J. Adler and D. I. Santiago, Mod. Phys. Lett. A14, 1371 (1999).

\bibitem{ZuWh} N. Bohr, "Discussions with Einstein
on Epistemological Problems in Atomic Physics", in {\em Quantum
Theory and Measurement}, edt. J. A. Wheeler and H. W. Zurek
(Princeton University Press, Princeton, 1983).

\bibitem{BR} N. Bohr and L. Rosenfeld, Kgl. Danske Videnskab S. Nat. Fys. Medd. 12,
1 (1933); N. Bohr and L. Rosenfeld, Phys. Rev. D78, 794 (1950);  A.
Peres and N. Rosen, Phys. Rev. 118, 335 (1960).

\bibitem{BaCa} P. M. V. B. Barone and A. O. Caldeira, Phys. Rev.
A43, 57 (1991).

\bibitem{HM2} B. L. Hu and A. Matacz,   Phys. Rev.
D49, 6612 (1994).

\bibitem{AnZ} C. Anastopoulos and A. Zoupas, Phys. Rev. D58, 105006
(1998).

\bibitem {JH} P. R. Johnson, Ph. D. thesis, University of Maryland, 1999.
P. R. Johnson and B. L. Hu, Phys. Rev. D65, 065015 (2002); Found.
Phys. 35,  1117 (2005) [gr-qc/0501029].

\bibitem{LH07} Shih-Yuin Lin and B. L. Hu, Phys. Rev. D 76, 064008
(2007).


\bibitem{timemuga} J. G. Muga, R. Sala Mayato and I. L. Egusquiza
(editors), {\em Time in Quantum Mechanics} (Springer, Berlin, 2002);
J. G. Muga and C. R. Leavens, Phys. Rep. 338, 353 (2000).


\bibitem{AnSav06} C. Anastopoulos and
N. Savvidou, J. Math. Phys. 47, 122106 (2006).


\bibitem{AnaHalUnc}
A. Anderson and J. J. Halliwell, Phys. Rev. D48, 2753 (1993). C.
Anastopoulos and J. J.  Halliwell, Phys. Rev. D51, 6870 (1995).


\bibitem{HuZhaUnc}
B. L. Hu and Y. Zhang, Mod. Phys. Lett. A8, 3575 (1993);
 Int. J. Mod. Phys. 10, 4537 (1995)


\bibitem{relational_time} C. Rovelli, Phys. Rev. D43, 442 (1991); M.
Montesinos, C. Rovelli and T. Thiemann,     Phys.Rev. D60, 044009
(1999); R. Gambini and R. A. Porto, Phys.Rev. D63, 105014 (2001).

\bibitem{critrel} J. B. Hartle, Class. Quant. Grav. 13, 361 (1996);
N. Savvidou and C. Anastopoulos, Class. Quant. Grav. 17, 2463
(2000).

\bibitem{IL} C. J. Isham, J. Math. Phys. {\bf 35},
2157 (1994); C. J. Isham and N. Linden, {\it ibid.} {\bf 35}, 5452
(1994).

\bibitem{ILSS} C. J. Isham and N. Linden, J. Math. Phys. {\bf 36}, 5392
(1994); C. J. Isham, N. Linden, K. Savvidou and S. Schreckenberg, J.
Math. Phys. {\bf 39}, 1818 (1998); K. Savvidou, J. Math. Phys. 40,
5657 (1999); gr-qc/9912076; Braz. J. Phys. 35, 307 (2005).

\bibitem{timeaver} J. B. Hartle,
Phys. Rev. D44, 3173 (1991); N. Yamada and S. Tagaki, Prog. Theor.
Phys. 85, 985 (1991); 86, 599 (1991); 87, 77 (1992); A. W. Bosse and
J. B. Hartle, Phys. Rev. A 72, 022105 (2005); D. Sokolovski, Phys.
Rev. A 57, R1469 (1998);  Phys. Rev. A 59, 1003 (1999); Y. Liu and
D. Sokolovski, Phys. Rev. A 63, 014102 (2001).

\bibitem{Ana06} C. Anastopoulos,  Ann. Phys. 313, 368 (2004); Found. Phys. 36, 1601 (2006).


\end{thebibliography}
\end{document}